\documentclass[smallextended]{svjour3}       
\smartqed  
\usepackage{graphicx,amsfonts}
\journalname{General Relativity and Gravitation}
\usepackage{amsfonts,amssymb,amsbsy}
\begin{document}

\title{Point-like source solutions in modified gravity\\
with a critical acceleration}

\author{Ja.V. Balitsky  \and V.V. Kiselev}


\institute{
           Ja.V. Balitsky \at
              Moscow Institute of Physics and Technology
              (State University), Russia, 141701, Moscow Region,
              Dolgoprudny, Institutsky 9
              \and
V.V.Kiselev \at
              Russian State Research Center
              Institute for High Energy Physics
              (National Research Centre Kurchatov Institute),
              Russia, 142281, Moscow Region, Protvino, Nauki 1\\
              Tel.: +7-4967-7138-63\\
              Fax: +7-4967-7428-24
              \at
              Moscow Institute of Physics and Technology
              (State University), Russia, 141701, Moscow Region,
              Dolgoprudny, Institutsky 9\\
              \email{Valery.Kiselev@ihep.ru}           
}

\date{Received: date / Accepted: date}

\maketitle

\begin{abstract}
We consider equations of modified gravity involving critical accelerations
and find its solutions for the point-like source by suggesting the
appropriate symmetry of metrics in the empty space-time.

\end{abstract}

\section{Introduction}
\label{intro}

The modified Newtonian dynamics \cite{MOND} is a successful phenomenological
setting for the description of ''dark matter effects'' at galactic scales
\cite{McGT}. Its concept is based on the introduction of a universal critical
acceleration $g_0$ as the manifestation of empirical regularities in dark
matter halos, so that the gravitation law is crucially modified at
accelerations less than $g_0$ with the ordinary visible or baryonic sources
in the following way
\begin{equation}\label{MOND}
    {\boldsymbol g}\,\zeta\left(\frac{g}{g_0}\right)=-\boldsymbol \nabla\phi_M,
\end{equation}
where the interpolating function $\zeta$ is originally set to be equal to
\begin{equation}\label{interpolate}
    \zeta(y)=\left(\frac{y^2}{
    {1+y^2}}\right)^\frac12,
\end{equation}
while $\phi_M$ is the gravitational potential of visible matter with density
$\rho_M$: $\triangle\phi_M=4\pi G\rho_M$.

In the Newtonian limit $y\gg 1$ of super-critical accelerations one finds
corrections negligible for the Solar system, for instance. In the deep
MONDian limit $y\ll 1$ of sub-critical accelerations the point-like source of
mass $M$ produces the enforced acceleration in the same direction as the
Newtonian one,
\begin{equation}\label{deep}
    g=\frac1r\,\sqrt{g_0GM}.
\end{equation}
This law reproduces flat rotation curves in dark galactic halos of disc
galaxies without introduction of any dark matter.

However, actually we cannot a priori fix a precise form of the function
$\zeta(y)$, since the main features of the galactic dynamics are predicted
with an impressive accuracy using only the asymptotic behavior of $\zeta(y)$:
$\zeta(y)\rightarrow1$ for $y\gg 1$ and $\zeta(y)\rightarrow y$ for $y\ll 1$.
So, there are some families of interpolating functions suitable for the same
predictions, say,
\begin{equation}
\zeta(y)\mapsto \zeta_n(y)=\frac{y}{(1+y^n)^{\frac{1}{n}}}
\end{equation}
The experimental data of galaxy disks \cite{Fit} prefer for n posed between 1
and 2. For a complicated $\gamma$-family \cite{FitSolarSystem}
\begin{equation}
\zeta(y)\mapsto \xi_\gamma(y)=\left(1-\exp\{-\sqrt{y^\gamma\zeta^\gamma(y)}\}\right)
^{-\frac{1}{\gamma}},
\end{equation}
combined data on galaxies and the Solar system can be well described at
$\gamma=1$, too. In this respect, we accept the form of (\ref{interpolate})
in order to reach the consistency with the data at the galaxy scale of
distances at the rather simple functional expression.

In addition to the asymptotic flatness of the rotation curves in spiral
galaxies, the most impressive successes of modification (1) are the baryonic
Tully-Fisher relation (BFTR), Faber-Jackson relation and Freeman limit
observed as the dynamical regularities in galaxies. Let us shortly mention
these regularities.

With a remarkably little intrinsic scatter, BTFR reads of
\begin{equation}
\log M_b = s \log V_f - \log b
\end{equation}
for $M_b=M_\ast+M_g$ being the sum of masses of visible stars and gas in the
disk galaxy, and $V_f$ being the flat limit of the rotation curve, while
$b=Gg_0$ as it directly follows from (\ref{deep}), and the slope $s=4$.
Moreover, the BTFR is a remarkably persistent relation, as it holds for both
low and high surface brightness galaxies.

The Faber-Jackson relation holds with quite a good accuracy and suggests that
for quasi-isothermal elliptical galaxies with dispersion of radial velocity
$\sigma$ and mass $M$
\begin{equation}
\sigma^4\sim GM g_0.
\end{equation}
This follows from the Jeans equation
\begin{equation}
\frac{d\sigma^2}{dr}+\sigma^2\frac{2\beta+\alpha}{r}=-g(r)
\end{equation}
at constant $\alpha=\frac{d\ln\rho}{d\ln r}$ and
$\beta=1-\frac{(\sigma^2_\theta+\sigma^2_\phi)}{2\sigma^2}=0$, so that
\begin{equation}
    \frac{d(\sigma^2 \rho)}{dr}=-\frac{\rho\sqrt{GMg_0}}{r}\quad
    \Rightarrow\quad
    \sigma^4=\frac{GMg_0}{\alpha^{2}}.
\end{equation}

Moreover, the Milgrom's law provides an additional stability for stellar
systems regardless of the form of interpolating function. This is known as
the Freeman limit: the disks with the surface density
$\Sigma\leqslant\frac{g_0}{G}$ have enhanced stability \cite{Stability} and
the number of disks with $\Sigma\geqslant\frac{g_0}{G}$ decreases
exponentially. So, since the bulk of the disk is in the weak-accelerated
regime, the acceleration $g\sim\sqrt{M}$, instead of $g\sim M$ in the
Newtonian regime. Thus, $\frac{\delta g}{g}=\frac{\delta M}{2M}$ instead of
$\frac{\delta g}{g}=\frac{\delta M}{M}$, leading to the weaker response to
the small perturbations. Numerical simulations show, that MOND has an effect,
similiar to the dark matter halo in stabilizing the galaxy disk
\cite{NumericalSimulation}.

So, the critical acceleration $g_0$ is observed phenomenological quantity,
and it is measured with high accuracy in the gas-reach galaxies \cite{g_0}

\begin{equation}
g_0=(1.21\pm0.14)\cdot10^{-10}\mbox{m s}^{-2}
\end{equation}





The generalization of nonrelativistic relation (\ref{MOND}) to the field
theory is given, for instance, in the Tensor-Vector-Scalar theory (TeVeS) by
J. Bekenstein \cite{JB}, wherein he introduces additional gravitational
vector and scalar fields replacing the dark components in the general
relativity (GR). Other models are discussed in \cite{McGT,FT-models}.
However, those models involve ad hoc new degrees of freedom. Moreover, they
have some deep conceptual problems like instabilities \cite{McGT}. Recently,
a nonlocal origin of modified gravity has been also discussed \cite{RW}. The
origin of critical acceleration has been considered in numerous articles
\cite{Pazy,Pazy2,Pazy3,Li,Kiselev,Klinkhamer,Neto}.

So, in order to investigate generic features of modifies gravity with the
critical acceleration, we prefer for a more phenomenological approach by
investigating the approximate equations and their generic consequences if
applicable. In this way, we can rewrite the law of modified gravity in
(\ref{MOND}) with interpolating function (\ref{interpolate}) in terms of
Ricci curvature tensor as
\begin{equation}\label{Ricci}
    \int d^3r R_0^0\cdot\zeta\left(\frac{
    \int d^3r R_0^0
    }{
    {
    \int d^3r K_0^0
    }}\right)
    =
    \int d^3r \bar R_0^0,
\end{equation}
where the bar tensor is defined in terms of energy-momentum tensor
$T_\mu^\nu$ for the visible matter,
\begin{equation}\label{matter-R}
    \bar R_\mu^\nu=8\pi\,G\left(T_\mu^\nu-\frac12\,\delta_\mu^\nu\,T\right),\qquad
    T=T_\mu^\mu,
\end{equation}
and the external curvature is given by
\begin{equation}\label{ext-K}
    K_0^0=g_0\,\frac2r,
\end{equation}
while the integration in (\ref{Ricci}) operates in the sphere of radius $r$.
Since in the non-relativistic limit $R_0^0\approx \triangle\phi$ at
$g_{00}\approx 1+2\phi$, we can easily reproduce the deep MONDian limit with
the ``dark matter mass'' $M_{DM}=\int d^3r\,\rho_{DM}$ entering
$\triangle\phi= 4\pi\,G(\rho_M+\rho_{DM})$, so that $GM_{DM}\approx r\sqrt{GM
g_0}$. The Newtonian limit is also restored. The main feature of
(\ref{Ricci}) is the presence of external Ricci tensor $K^\mu_\nu$ as the
critical divider of two regimes.

Note that the equation (\ref{Ricci}) takes into the account the symmetry of
space-time: the rotations and stationarity. It does not represent the full
system of equations for a modified gravity with the critical acceleration,
but it gives the relativistic generalization of MONDian dynamics in the
symmetric case. In addition to (\ref{Ricci}) we have to require the
conservation law for the energy-momentum tensor of matter, of course.
Nevertheless, for instance, we can get the relativistic static solution in
the modified gravity with the critical acceleration in the case of point-like
source, that conserve both the spherical symmetry and stationarity. This
symmetry in general relativity provides the metrics of the form
\begin{equation}\label{metrics}
    ds^2=f(r)\,dt^2-\frac{1}{f(r)}\,dr^2-r^2(d\theta^2+\sin^2\theta d\phi^2),
\end{equation}
due to the symmetry of source energy-momentum tensor: $T_0^0=T_r^r$
\cite{CQG-K}. It is natural to extend this symmetry of metrics to the case of
modified gravity with the point-like source to find the function $f(r)$.

It is spectacular that (\ref{Ricci}) can be extrapolated in cosmology with a
little modification in order to conserve the spatial homogeneity at
extra-galactic scales in large scale structre of Universe as it was done in
\cite{CosExMOND}: one can simply replace the external Ricci curvature by de
Sitter one, $K_0^0\mapsto \bar K_0^0=3g_0'$, while the integration inside the
sphere is canceled, that gives
\begin{equation}\label{Ricci-CosEx}
    R_0^0\cdot\zeta\left(\frac{R_0^0}{
    \bar K_0^0}
    \right)=\bar R_0^0.
\end{equation}

In this paper we find the point-like source solutions of (\ref{Ricci}) in the
symmetric case of (\ref{metrics}) for both external Ricci tensors of $K_0^0$
and $\bar K_0^0$.

\section{The modified solutions}
In the symmetric case of (\ref{metrics}), when $R_0^0=\triangle f/2$, the
integrating over the volume of finite sphere with radius $r$ yields
\begin{equation}
    \int d^3 r R ^0_0=
    \frac12\int d\boldsymbol S \cdot\nabla f(r)=
    \frac12\int r^2d\Omega\,\frac{\boldsymbol r}{r}\,
    \frac{df}{dr}\,\frac{\boldsymbol r}{r}=
    2\pi r^2 \frac{df}{dr}.
\end{equation}
Therefore
\begin{equation}\label{Ricci2}
    \left(\int d^3 r \bar{R} ^0_0\right)^2=\frac{\displaystyle
    \left(2\pi r^2\frac{df}{dr}\right)^4}{
    {\displaystyle
    \left(2\pi r^2\frac{df}{dr}\right)^2+\left(\int d^3r K_0^0\right)^2}},
\end{equation}
hence, the general solution of (\ref{Ricci2}) is
\begin{equation}
    \frac{df}{dr}=\frac{\int d^3 r \bar{R} ^0_0}{2\pi r^2\sqrt{2}}
    \left\{1+\sqrt{1+\left(\frac{2\int d^3 r K^0_0}
    {\int d^3 r \bar{R} ^0_0}\right)^2}\right\}^\frac12.
\end{equation}
At $K_0^0\equiv 0$, we reproduce the limit of general relativity, of course.
However, complete equations of GR allow us to get the more definite
expression with the same condition of symmetry,
\begin{equation}
    \frac{df}{dr}=-\frac{d}{dr}\frac{2G M_\mathrm{tot}(r)}{r},
\end{equation}
at $M_\mathrm{tot}(r)=4\pi\int dr\,r^2\rho(r)$ being the total mass enclosed
in the sphere of radius $r$.

For MOND we get
$$
    \int d^3 r\,K_0^0=4\pi\,g_0 r^2,
$$
while for the case of de Sitter external Ricci tensor we find
$$
    \int d^3 r\,\bar K_0^0= 4\pi\,g_0' r^3.
$$
So, the solutions in MOND and its modification derived at cosmic scales
should differ essentially.

\subsection{De Sitter vacuum state}
Considering a solution of modified gravity, we can analyze basic physical
features of changing. So, the vacuum with the positive density of energy
$\rho_\Lambda$ corresponds to de Sitter space. Then,
$$
    \bar R_0^0=-8\pi\,G\rho_\Lambda,
$$
yielding
$$
    \int d^3 r\,\bar R_0^0=-\frac{32\pi^2}{3}\,G\rho_\Lambda r^3.
$$
Therefore, MOND breaks down the vacuum homogeneity, since
\begin{equation}
    \frac{df}{dr}^\mathrm{MOND}=-\frac{16\pi}{3}\rho_\Lambda G r
    \,\frac{1}{\sqrt{2}}\,
    \left\{1+\sqrt{1+\left(\frac{3g_0}
    {4\pi G \rho_\Lambda r}\right)^2}\right\}^\frac12,
\end{equation}
and the vacuum metrics,
$$
    f^\mathrm{dS}=1-\frac{8\pi}{3}\,G\rho_\Lambda r^2
$$
transformed into the solution, that can be treated in the framework of
general relativity at $r\to 0$ as the metric given by a singular negative
density of matter $\rho_m\sim -1/\sqrt{r}$. This behavior artificially
introduces a point of world center.

In contrast, when the external Ricci tensor is defined by de Sitter space,
too, i.e. $\bar K_0^0=3 g_0'=-8\pi\,G\tilde\rho_\Lambda$, the modified metric
scales like the metric of de Sitter space at the modified density,
\begin{equation}
    \frac{df}{dr}^\mathrm{mdS}=-\frac{16\pi}{3\sqrt{2}}\,\rho_\Lambda G r
    \left\{
    {1+\sqrt{1+
    \left(\frac{\tilde\rho_\Lambda}{\rho_\Lambda}\right)^2}}
    \right\}^\frac12,
\end{equation}
i.e.
$$
    \rho_\Lambda\mapsto \rho_\Lambda'=\frac{\rho_\Lambda}{\sqrt{2}}
    \left\{1+\sqrt{1+\left(\frac{\tilde\rho_\Lambda}
    {\rho_\Lambda}\right)^2}\right\}^\frac12.
$$
It is interesting to note, that there is the solution with
$\rho_\Lambda'=\tilde\rho_\Lambda$, if $\tilde\rho_\Lambda=x \rho_\Lambda$,
while $x$ satisfies the following relation:
\begin{equation}
    \sqrt{1+\sqrt{1+x}}=\sqrt{2}x,
\end{equation}
that gives
\begin{equation}
    x=\frac{\sqrt{5}}{2}\approx1.12.
\end{equation}

It is important to stress that the modified gravity with the scaled critical
acceleration conserves the vacuum, i.e.  de Sitter space remains de Sitter
space. This fact is important for the cosmological extrapolation of MOND,
since the cosmic evolution starts with the description of homogeneous
Universe.

\subsection{Point-like source}
The black hole with mass $M$ corresponds to
\begin{equation}
    \int d^3r \bar R^0_0=4\pi G M.
\end{equation}
Therefore, the MONDian solution is given by
\begin{equation}\label{M-MOND}
    \frac{df}{dr}^\mathrm{MOND}=\frac{2GM}{r^2}\,\frac{1}{\sqrt{2}}\,
    \left\{1+\sqrt{1+\left(\frac{2g_0 r^2}{GM}\right)^2}\right\}^\frac12,
\end{equation}
while the modification with the scaled critical acceleration results in
\begin{equation}\label{M-ext}
    \frac{df}{dr}^\mathrm{mdS}=\frac{2GM}{r^2}\,\frac{1}{\sqrt{2}}\,
    \left\{1+\sqrt{1+\left(\frac{2g_0^\prime r^3}{GM}\right)^2}\right\}^\frac12.
\end{equation}
So, at $g_0^\prime r_S=g_0$ solutions (\ref{M-MOND}) and (\ref{M-ext}) can be
matched at $r=r_S$.

We expect that effects of modified gravity are essential at galactic scales.
In order to test this expectation in the Solar system, we estimate the
corrections to the magnitude of the acceleration of Neptune
$g=|\frac{1}{2}\frac{df}{dr}|$ in both cases of modification, because the
Neptune is the most distant planet from the Sun that has an almost circular
orbit, since
\begin{equation}
    \frac{\mbox{Aphelion-Perihelion}}{\mbox{Aphelion}}\approx 0.02.
\end{equation}
Expanding at $\frac{g r^2}{GM_\odot}\ll 1$ in the MOND case, we find
\begin{equation}
    \frac{d f}{d r}^\mathrm{MOND}
    \approx \frac{2GM_\odot}{r^2}\left\{1+
    \frac{1}{2}\left(\frac{g_0 r^2}{GM_\odot}\right)^2\right\}=
    \frac{r_g}{r^2}\left\{1+2\left(\frac{g_0 r^2}{r_g}\right)^2\right\},
\end{equation}
where $r_g=2GM_\odot$ is the gravitational radius of the Sun. Substituting
the semi-major axis of Neptune $r\mapsto a\approx 4.5\cdot10^9$ km, we get
\begin{equation}
    \frac{\bigtriangleup g}{g}\approx 2 \left(\frac{g_0 r^2}{r_g }\right)^2
    \approx1.6\cdot 10^{-10}.
\end{equation}
The limit of deep MOND ($\frac{g_0 r^2}{GM}\gg1)$ of flat rotation curves
with $g=V_f^2/r$ is also recovered
\begin{equation}
    \frac{df}{d r}^\mathrm{MOND}\approx 2g\approx
    \frac{2}{r}\sqrt{GMg_0}.
\end{equation}


In the case of scaled critical acceleration the limit
$\frac{2g_0'r^3}{GM_\odot}\ll1$ yields
\begin{equation}
    \frac{df}{dr}^\mathrm{mdS}\approx\frac{2GM_\odot}{r^2 }\left\{1+\frac{1}{2}
    \left(\frac{r^3 g_0^\prime}{GM_\odot }\right)^2\right\},
\end{equation}
and the correction to the gravitational acceleration is equal to
\begin{equation}
    \frac{\bigtriangleup g}{g}\approx
    \frac{1}{2}
    \left(\frac{r^3 g_0^\prime}{GM_\odot }\right)^2=
    \frac{32 \pi^2}{9}\left(\frac{r^3 \rho_{\Lambda}}{M_\odot}
    \right)^2\approx 3\cdot 10^{-36},
\end{equation}
where we estimate the slope of critical acceleration by
$g_0'=8\pi\,G\rho_\Lambda/3$ at the observed value of cosmological constant
$\rho_\Lambda$.

At $\frac{2g_0'r^3}{GM}\gg1$ we obtain
\begin{equation}
    \frac{df}{dr}^\mathrm{mdS}\approx 2g\approx
    2\sqrt{\frac{GMg_0'}{r}},
\end{equation}
that covers the limit of $g_0=g_0'r$ again in agreement with the
consideration in cosmology \cite{CosExMOND}.

\subsection{Black hole surrounded by quintessence}

The point-like source of black hole can be generalized to the case of
symmetric metric (\ref{metrics}) by adding some sources with $T_0^0=T_r^r$
and constant equation of state parameter $w_q$ \cite{Quintessence}, when in
general relativity
$$
    f=1-\frac{2GM}{r}+\sum_q \left(\frac{r_q}{r}\right)^{3w_q+1},
$$
so that
\begin{equation}
    \int d^3r \bar R^0_0=4\pi GM 
    -2\pi \sum_q r_q (3w_q+1)\left(\frac{r_q}{r}\right)^{3w_q},
\end{equation}
where $r_q$ is the quintessence parameter of length dimension.

For instance, the charged black hole corresponds to the quintessence of
static electromagnetic field with $w_q\mapsto \frac13$ and $r_q^2=GQ^2$, so
that
\begin{equation}
    \int d^3r R^0_0=4\pi GM -\frac{4\pi G Q^2}{r}=4\pi G M A(r),\qquad
    A(r)=\left(1-\frac{Q^2}{Mr}\right).
\end{equation}
Then, we find the modifications of Reissner--Nordstr{\o}m solutions
\begin{equation}\label{RN-MOND}
    \frac{df}{dr}^\mathrm{MOND}=\frac{2GM}{r^2}\,
        \frac{A(r)}{\sqrt{2}}\,
    \left\{1+\sqrt{1+\left(\frac{2g_0 r^2}{GM A(r)}\right)^2}\right\}^\frac12,
\end{equation}
and
\begin{equation}\label{RN-ext}
    \frac{df}{dr}^\mathrm{mdS}=\frac{2GM}{r^2}\,\frac{A(r)}{\sqrt{2}}\,
    \left\{1+\sqrt{1+\left(\frac{2g_0^\prime r^3}{GM A(r)}\right)^2}
    \right\}^\frac12.
\end{equation}
At large distances
$$
    \frac{df}{dr}^\mathrm{MOND}\approx 2\frac{\sqrt{GMg_0}}{r}\left(1-
    \frac{Q^2}{2Mr}\right),\quad
    \frac{df}{dr}^\mathrm{mdS}\approx 2\sqrt{\frac{GMg_0'}{r}}\left(1-
    \frac{Q^2}{2Mr}\right).
$$

At $w_q=-\frac13$ and $M=0$ we recover the case of global monopole
\cite{GlobalMonopole}, when $f=1-\kappa$ and it is not changed by the
modification with the critical acceleration.

\section{Conclusion}

We have considered solutions of modified gravity with critical acceleration
for the case of symmetry in the metric as it is relevant to the symmetry in
the general relativity for the gravitational sources.

First, we note, that homogeneous distribution of matter in the de Sitter
space is not consistent with the original MOND, while the critical
acceleration scaled with distance conserves the de Sitter structure of space.
This fact does not mean that the MOND is incorrect, it only points to that
MOND has got a restricted area of applicability, i.e. it well works for
inhomogeneous distributions of matter of island kind. Moreover, it could mean
that there is a regulator of regimes in the modified gravity with the
critical acceleration, that would be an extra field inherently related to the
gravity.

Second, we have found the metrics for point-like sources of gravity in the
modified models of gravity, i.e. the modified Schwarzschild black holes and
Reissner--Nordstr\o m black holes with the electric charge. The global
monopole solution is not changed by the gravity with the critical
acceleration. We have compared corrections to the gravitational acceleration
in far regions of Solar system, for example, at the distance of Neptune, so
that gravity with the scaled critical acceleration has got a much smaller
correction.

In order to study more general solutions of modified gravity with the
critical acceleration, we need a complete set of relevant gravitational
equations, that suggests the involvement of models to the moment.

Next, conditions of transition from the fixed critical acceleration to the
scaled one require further investigations. Moreover, one could expect that
Newtonian dynamics might be restored at extremely small accelerations
\cite{Newscale,epsilon}.

This work is supported by  Russian Scientific Foundation, project \#
14-12-00232.

\end{document}